\documentclass[aps,prd,reprint,superscriptaddress,nofootinbib]{revtex4-1}
\usepackage{graphicx}
\usepackage{xcolor}
\graphicspath{ {images/} }
\usepackage{dcolumn}
\usepackage{ytableau} 
\usepackage{slashed} 
\usepackage[enableskew]{youngtab} 
\usepackage{braket}
\usepackage{harpoon}
\usepackage{mathtools}
\usepackage[
	colorlinks=true, 
    filecolor=black,      
    urlcolor=black,
    linkcolor=black,
    citecolor=red]
						{hyperref}
\usepackage{cleveref}

\usepackage{mathtools}
\usepackage{tikz}
\usetikzlibrary{calc,decorations.pathreplacing,calligraphy,matrix}
\graphicspath{{./Figs/}}
\tikzset{Dotted/.style={
    line width=\pgfkeysvalueof{/tikz/Young/dot size},
    dash pattern=on 0.001\pgflinewidth off #1,line cap=round,
    shorten <=#1},Dotted/.default=3pt,
    vdots/.style={draw=none,path picture={
     \draw let \p1=(path picture bounding box.north),
        \p2=(path picture bounding box.south) in
        [Dotted={(\y1-\y2)/4}]
        (\p1) -- (\p2);
    }},
    cdots/.style={draw=none,path picture={
     \draw let \p1=(path picture bounding box.east),
        \p2=(path picture bounding box.west) in
        [Dotted={(\x1-\x2)/4}]
        (\p2) -- (\p1);
    }},
    Young tableau/.style={matrix of math nodes,nodes in empty cells,
    nodes={draw,minimum size=\pgfkeysvalueof{/tikz/Young/cell size},inner sep=0.5pt},
    column sep=-\pgflinewidth,row sep=-\pgflinewidth},
    Young/.cd,cell size/.initial=1.75em,
    dot size/.initial=1.2pt
    }
\begin{document}
\title{Chiral symmetry breaking and confinement:  separating the scales}
\author{Nick Evans}
\email{n.j.evans@soton.ac.uk}
\affiliation{School  of  Physics  \&  Astronomy  and  STAG  Research  Centre,  University  of  Southampton, \\
Highfield, Southampton  SO$17$ $1$BJ, UK.\\
}
\author{Konstantinos S.~Rigatos}
\email{k.c.rigatos@soton.ac.uk}
\affiliation{School  of  Physics  \&  Astronomy  and  STAG  Research  Centre,  University  of  Southampton, \\
Highfield, Southampton  SO$17$ $1$BJ, UK.\\
}
\begin{abstract}
\noindent We review arguments that chiral symmetry breaking is triggered when the quark bilinear condensate's dimension passes through one ($\gamma=1$). This is supported by gap equations and more recently holographic models. Confinement may then be a separate property of the pure Yang-Mills theory below the scale of the dynamically generated quark mass, occurring at the scale of the pole in the deep IR running. Here, we use perturbative results for the running of the gauge coupling and $\gamma$ in asymptotically free SU($N$) gauge theories with matter in higher dimension representations to seek the best candidate theories where confinement and chiral symmetry breaking can be maximally separated. For example, SU(2) gauge theory with a single Weyl quark in the $S_3$ (dimension 4) representation may have a factor of 20 separation in scale. Such a theory could be simulated on the lattice to test the separation. We also propose studying multi-representation theories where the higher dimension representation forms a condensate at one scale that can be quite separate from the condensation scale of the second representation matter.  The confinement scale would presumably be below the second scale. For example, SU(3) gauge theory with a Weyl adjoint fermion and ten fundamental quarks may have a separation of a factor of 20 also. 
\end{abstract}
\maketitle
\section{Introduction}

QCD is an asymptotically free $SU(3)$ gauge theory with fermions in the fundamental representation. The coupling becomes strong in the infrared at around $200-300$ MeV. Near this scale, the dynamics triggers chiral symmetry breaking by the dynamical generation of a quark mass or a vacuum expectation for the operator $\bar{q}_L q_R + h.c$ \cite{Weinberg:1975gm}. Experimentally the signatures of this symmetry breaking are the pseudo-Goldstone nature of the pions and the absence of parity doubling in the spectrum. In addition, one observes colour charge confinement - that is the quarks cannot be liberated from hadrons (although the pair creation of light quarks breaks flux tubes between the charges if they are too separated). Here it has been posited that the mechanism is the condensation at strong coupling of magnetically charged scalar operators leading to a dual Meissner effect \cite{thooft,Mandelstam:1978ed}. 

It has long been a question of interest as to whether or not chiral symmetry breaking and confinement are separate dynamics or inherently linked in QCD - see for example \cite{Casher:1979vw,Shuryak:1981fz,Gross:1991pk,Bak:2004nt, Synatschke:2008yt,Lang:2011vw,Glozman:2012iu,Suganuma:2017syi,Yang:2020hun}. A common back of the envelope computation is to use the confinement distance in Heisenberg's uncertainty principle to estimate the momentum of the state and hence its energy and mass. The logic here is that only confinement would need chiral symmetry breaking and not vice versa. Of course, the masslessness of the pions in the chiral limit provides a (symmetry driven) counter example to such simple logic. That the two phenomena are linked is supported by lattice simulations of QCD \cite{Aoki:2006br,Karsch:2001cy} which show a single finite temperature phase transition at which the theory both deconfines and restores chiral symmetry. On the other hand, that transition is a blurred cross-over and, further, lattice simulations of QCD with Nambu-Jona-Lasinio operators present have already shown that a clean split between the chiral symmetry breaking scale and the confinement scale can be realized \cite{Sinclair:2008du} - the higher dimension operator raises the chiral symmetry breaking scale. Here we will pursue the idea that the scales are separate, but somewhat accidentally coincide for fundamental representation quarks. We will explore asymptotically free gauge theories, without higher dimension operators, but with fermions in different representations than the fundamental to seek for cases where the separation can be maximised. We will search for theories where this separation of scales could be tested with first principle lattice simulations (and review simulations already done that support these ideas).

In the 1980s chiral symmetry breaking was studied using gap equation techniques \cite{Higashijima:1983gx,Miransky:1984ef,Appelquist:1986an}. That is truncations of the Schwinger-Dyson equation for the quark self energy were studied in the rainbow approximation and the resulting equations were shown to have a critical coupling for chiral symmetry breaking. Additionally, it was understood that the key criteria for mass generation in the calculation was that the dimension of the quark bilinear $\gamma$ grew greater than one - see for example \cite{Cohen:1988sq}. This is a natural point for an instability to set in since it is when the quark mass and condensate become equal in dimension. These methods neglected confinement. The justification is as follows: the assumption is that the QCD coupling grows until the $\gamma=1$ criteria is met, at which point a dynamical quark mass forms.  The quarks should then be integrated from the theory leaving a pure Yang-Mills theory that will run to yet stronger coupling. Confinement is then imagined to be a product of that pure glue theory, which perhaps contains condensed magnetic monopole configurations. For quarks in the fundamental representation the value of the critical coupling, $\alpha_c$, when $\gamma=1$ is strong and so the pure Yang-Mills theory starts at rather large coupling and runs very fast, so its Landau pole and confinement lie rather close to the chiral symmetry breaking scale. The assumption of this  separation was necessary to justify the use of gap equation methods but not proven. The early gap equation work has been extended since to include the Schwinger Dyson equation for the gluon propagator, where some hints of signals of confinement are possible (see for example \cite{Hopfer:2014zna}), but the basic picture does not seem to change. 

More recently holography \cite{Maldacena:1997re,Witten:1998qj,Gubser:1998bc} has provided further insights into chiral symmetry breaking \cite{Babington:2003vm}.  Here the quark bilinear is mapped to a scalar in an AdS$_5$ like space where the fifth dimension of AdS, $r$,  corresponds to energy scale in the gauge theory. The symmetry breaking becomes a Higgs like mechanism where the scalar condenses at small $r$ because its mass is driven through the Breitenlohner-Freedman (BF) bound \cite{Breitenlohner:1982jf} so that the zero vacuum expectation value configuration is unstable. In AdS$_5$/CFT$_4$, the mass of the scalar is related to the dimension, $\Delta$  of the dual operator by $m^2=\Delta(\Delta-4)$. The case $\Delta=2$ (again $\gamma=1$) corresponds to the saturation of the BF bound. A controlled example of such a chiral symmetry breaking theory is provided by the D3/probe-D7 system with a magnetic field on the probe brane \cite{Filev:2007gb, Erdmenger:2007bn}. The base gauge theory is a ${\cal N}=4$ glue theory with $N_f$ quenched ${\cal N}=2$ hypermultiplets. The magnetic field breaks supersymmetry and conformal symmetry and triggers the formation of a chiral condensate. Analysis of the chirally symmetric theory shows precisely the mechanism where $m^2$ for the scalar is driven through the BF bound \cite{Alvares:2012kr}. The ${\cal N}=4$  glue of this theory is conformal and so there is no confinement even at infinitesimally small temperatures. The reader can find further examples of this holographic logic in \cite{Jarvinen:2011qe,Kutasov:2011fr,Alho:2013dka}. The key point here though is that these models support the gap equation criteria of $\gamma=1$ and that chiral symmetry breaking can be separate from confinement.

Let us therefore at least entertain that confinement and chiral symmetry breaking are separate. To shed further light on this phenomenon it would be good to build a bank of models where the separation is clear cut. We recognise that any number of such models will not prove the absence of a link in theories like QCD where the scales for both phenomena are so close but they will generate weight for the separation hypothesis. How might one find theories with such a separation? It was suggested a long time ago that theories with quarks in higher dimension representations may have higher values of $\gamma$ for a given  $\alpha$ and this would tend to raise the scale of the chiral symmetry breaking. In \cite{Marciano:1980zf}, it was ambitiously questioned whether a higher dimension representation quark in QCD might even provide a separation in symmetry breaking scales of 100 to allow QCD to generate the electroweak scale. The separations in scales are not expected to be this big as we will see but the philosophy remains. An additional problem that is likely to occur is that, as pointed out in \cite{Mocsy:2003qw,Mocsy:2004db}, the operators that cause confinement (the Polyakov loop expectation value) and chiral symmetry breaking (the quark condensate expectation value) will interact and if one forms at strong coupling then it may trigger the formation of the other. To break this link one would need the chiral symmetry breaking to happen at sufficiently weak coupling that the interactions between the two sectors remains small - we must seek theories with quite large separations in scale therefore. 

To make proposals for theories with separated scales we will perform here two very simple analyses based on perturbative results for the running of the coupling and $\gamma$. Our goal, in the spirit of \cite{Appelquist:1996dq, Ryttov:2007cx} for studies of the possibility of conformal window regimes, is to place down a ``straw man'' plot of the behaviour of gauge theories that lattice studies can be compared to, or motivated by. We will initially take the one loop results for the running of the gauge coupling and anomalous dimension and determine the value of $\alpha_c$ for chiral symmetry breaking in the space of asymptotically free SU$(N)$ theories. One can associate this value of the coupling with the scale of the quark mass or chiral symmetry breaking, $\Lambda_{\chi SB}$. We then set that value of $\alpha_c$ as the UV boundary condition on the pure Yang Mills theory below the quark mass scale. We simply compute the ratio of the Landau pole, as a measure of the confinement scale, to $\Lambda_{\chi SB}$. We also use the two loop runnings that include infra-red conformal fixed points to exclude theories where the infra-red (IR) coupling lies below the critical value - these theories lie in the conformal window. 

The majority of chiral symmetry breaking theories with a single representation have a predicted gap between $\Lambda_{\chi SB}$ and the confinement scale of 5 or less and at the level of this analysis it is hard to be certain whether they will in fact have a large gap. We will though highlight a couple of theories with high dimension representations that have a gap above ten which is of potential interest. We will review lattice work on this space of theories to see the degree to which the simple perturbative based results hold non-perturbatively.

Secondly we will look at multi-representation theories. A scenario one might dream of is a theory where the dynamics runs to strong enough coupling that one representation condenses and is integrated out, leaving a theory in the conformal window. The putative IR conformal theory might then never confine. Unfortunately this scenario does not seem to exist at the level of our approximations and one can not generate an infinite separation in scales between the mass gap of the higher dimension representation and the confinement scale. However, what one can do is to find theories with large gaps between the scale of the higher dimensional representation's chiral symmetry breaking scale and that for the lower dimensional representation.  The confinement scale should lie below the lower of these two scales. The most flexible case is to study theories with fundamental fermions and a minimum number of spinors of a higher  dimensional representation. By tuning the number of fundamental fermions the IR running of the coupling can be made to run more slowly than if the theory were pure glue (this is ``walking'' behaviour \cite{Holdom:1981rm}). We find a considerable number of examples with gaps between the two representations' chiral symmetry breaking scales greater than ten, and some where it is are as large as twenty to thirty. Of course it is possible, if our estimates are imperfect, that some of these theories truly enter the conformal window in the IR and the gap is much bigger. There has already been a small amount of lattice work on two representation theories which we review in this light.

Note that amongst the theories we consider are chiral theories with just a single Weyl  fermion in some real representation. Such theories are a challenge for the lattice since they potentially have a sign problem. However, there has been work on such theories where a Majorana mass is included and then it is experimentally tested in the simulations whether there is in fact a sign problem as the mass is reduced - see \cite{Bergner:2017gzw,Piemonte:2020wkm} for example. Progress has been made in this way and it may be possible by a judicious combination of continuum limit and chiral limit to avoid the sign problem completely (note that to identify the chiral symmetry breaking scale it is only necessary for the quark mass to lie at a lower scale, not to be formally zero). We include these theories therefore. Amongst our results though are plenty where Dirac representations are all that are needed to see gaps between confinement and chiral symmetry breaking.

The theories we propose with large gaps between the chiral symmetry breaking and confinement scales can in principle teach us about many aspects of strong coupling dynamics. Not only whether the two phenomena are truly separate but also, for example, how fermions decouple at strong coupling and how well the two loop running describes the non-perturbative walking regime of the gauge theories? We hope this work will inspire lattice studies of the phenomena we uncover. 

\section{Matter in one representation}

Asymptotic freedom in gauge theories was first computed at one loop in \cite{Gross:1973id,Politzer:1973fx} giving the classic result for the gauge coupling $\alpha$
\begin{equation}  \label{eq: beta_function1}
\begin{split}
\mu \frac{d \alpha}{d \mu} &= - b_0  \alpha^2, \\ 
b_0 &= \frac{1}{2 \pi} \left({11 \over 3} C_{2}(G) - {4 \over 3} \sum_{R} T(R)N_f(R) \right). 
\end{split}
\end{equation}
The running of the anomalous dimension for the quark bilinear operator is  given by
\begin{align} \label{eq: anomalous_dimension}
\gamma = \frac{3~C_2(R)}{2 \pi}~\alpha ~.
\end{align}
In these equations $T$ is half the Dynkin index, $C_2$ is the quadratic Casimir, and $N_f(R)$ is the number of flavours of the representation. The values of these constants (and the dimension of the representation $d(R)$) are given for the representations we will consider for $SU(N)$ theories in \cite{Feger:2012bs} and we reproduce them in \Cref{SUN_table}. The normalization for these group invariants that we choose is such that the Dynkin index in the fundamental representation is equal to $1$. The maximum number of flavours of any representation for the theory to be asymptotically free is controlled by requiring $b_0>0$. 

In this paper we will consider fermion representations that give asymptotically free theories for some choice of the number of flavours $N_f(R)$. Representations can be specified by their Dynkin indices or Young tableaux. The Dynkin indices of the singlet is just $(0~0~0~\cdots~0~0)$ and the Young diagram is $\bullet$.
The fundamental representation is $F=(1~0~0~\cdots~0~0)$ and the Young tableaux {\tiny $\vcenter{\hbox{\begin{tikzpicture}
   \node[Young tableau](yt1){ \\};
 \end{tikzpicture}}}$}. The remaining representations we consider are:\bigskip

\begin{tabular}{ccc}
Rank-n anti-&   Adjoint (G) & Rank-n\\
symmetric ($A_n$)&  & symmetric ($S_n$) \\
 $ \left. \right. \hspace{-1.cm}
 \qquad
 \vcenter{\hbox{\begin{tikzpicture}
   \node[Young tableau](yt4){ \\ 
   \\ 
   |[vdots]| \\
   \\
   \\};
  \draw[thick,decoration={calligraphic brace, amplitude=6pt},decorate] 
  (yt4-1-1.north east) -- (yt4-5-1.south east)
  node[midway,right=1ex]{$n$};
 \end{tikzpicture}}}$&  $ \left. \right.$ \hspace{-1cm}
 $
 \vcenter{\hbox{\begin{tikzpicture}
  \node[Young tableau](yt2){ & \\
   \\
   \\
   |[vdots]| \\
   \\};
  \draw[thick,decoration={calligraphic brace, amplitude=6pt},decorate] 
  (yt2-5-1.south west) -- (yt2-2-1.north west)
  node[midway,above=1ex,sloped]{$N-2$};
 \end{tikzpicture}}}$ &
$
 \vcenter{\hbox{\begin{tikzpicture}
   \node[Young tableau](yt3){ & & |[cdots]| & & \\};
  \draw[thick,decoration={calligraphic brace, amplitude=6pt},decorate] 
  (yt3-1-5.south east) -- (yt3-1-1.south west)
  node[midway,below=1ex]{$n$};
 \end{tikzpicture}}}  $  \\
 $(n~0~0~\cdots~0~0)$  &   $(1~0~0~\cdots~0~1)$ & $(0~0~0~\cdots~1~\cdots~0~0)$
 \end{tabular}

 \bigskip \bigskip \bigskip

 \begin{tabular}{ccc}
 $R_1$ & $R_2$ & $R_3$ \\
$ \vcenter{\hbox{\begin{tikzpicture}
   \node[Young tableau](yt5){ & \\ & \\ };
 \end{tikzpicture}}} $ & 
 $
 \vcenter{\hbox{\begin{tikzpicture}
   \node[Young tableau](yt6){ & \\  \\ };
 \end{tikzpicture}}}$  & $
 \qquad
 \vcenter{\hbox{\begin{tikzpicture}
   \node[Young tableau](yt7){ & \\
   \\
   \\
   |[vdots]| \\
   \\};
  \draw[thick,decoration={calligraphic brace, amplitude=6pt},decorate] 
  (yt7-5-1.south west) -- (yt7-2-1.north west)
  node[midway,above=1ex,sloped]{$N-3$};
 \end{tikzpicture}}}$  \\
  $(0~2~0~\cdots~0)$, &$(1~1~0~\cdots~0~0)$ &  $(1~0~\cdots~0~1~0)$\end{tabular} \bigskip
  
Theories with fermions in  these representations only are asymptotically free.

\begin{table*} 
\begin{tabular}{ |m{0.6cm} | m{3.5cm}| m{2.5cm}| p{3cm}|  m{3.5cm}|}  
 \hline
  R  & T(R) & C$_2$(R) & d(R) &N(R)$_{\rm max}$ \\[3mm]
\hline
 F & $\frac{1}{2}$ & $\frac{N^2-1}{2 N}$ & $N$ &$\frac{11 N}{2}$ \\[3mm]
  \hline
  G & $N$ & $N$ & $N^2-1$ & $\frac{11}{4}$\\[3mm]
\hline
$S_n$ & $\frac{(N-1)(N+n)(N+n-1)!}{2N(N^2-1)(n-1)!(N-1)!}$ & $\frac{n(N-1)(N+n)}{2N}$ & $\frac{(N+n-1)!}{n!(N-1)!}$ &$\frac{11 N^2(N+1)(n-1)!(N-1)!}{2(N+n)(N+n-1)!}$\\[3mm]
 \hline
$A_n$ & $\frac{(N+1)(N-n)(N-1)!}{2(N^2-1)(N-n)!(n-1)!}$ & $\frac{n(N-n)(N+1)}{2N}$ & $\frac{N!}{n!(N-n)!}$ & $\frac{11N(n-1)!(N-n)!}{2(N-n)(N-2)!}$ \\[3mm]
\hline
$R_1$ & $\frac{N^2-4}{6}$ & $\frac{2(N^2-4)}{N}$ & $\frac{(N-1)N^2(N+1)}{12}$ &$\frac{33N}{2(N^2-4)}$ \\[3mm]
\hline
$R_2$ & $\frac{N^2-3}{2}$ & $\frac{3(N^2-3)}{2N}$ & $\frac{(N-1)N(N+1)}{3}$ &$\frac{11N}{2(N^2-3)}$\\[3mm]
\hline
$R_3$ & $\frac{(N-2)(3N+1)}{2}$ & $\frac{(N-1)(3N+1)}{2N}$ & $\frac{(N-2)N(N+1)}{2}$ &$\frac{11N}{2(N-2)(3N+1)}$   \\[3mm]
\hline
\end{tabular}
\caption{Group theory factors for $SU(N)$ gauge theory representations and the maximum number of flavours for asymptotic freedom to be present at one loop. For SU(2) only the $F,G$ and $S_n$ represntations exist (here $G=S_2$). For SU(3) the $A_2=F$, $G=R_2$ and $R_3=S_2$. For SU(4) the $R_2$ becomes distinct and $R_3=R_2$. For $N>5$ all the representations are distinct.}
\label{SUN_table}
\end{table*}

\subsubsection{${\cal R} = \Lambda_{\rm pole}/\Lambda_{\chi SB}$}

To progress we must now specify a criteria for chiral symmetry breaking. We follow the logic of the papers \cite{Appelquist:1996dq, Ryttov:2007cx,Alho:2013dka}. For example, in the holographic models if one relates the mass squared of the scalar in AdS$_5$ to the dimension of the operator in the perturbative regime one has
\begin{equation} \begin{array}{ccl}
m^2 & = &   \Delta(\Delta-4) \\
 &=&(3 - \gamma)(-\gamma-1)\\
&  \simeq & -3 - 2 \gamma. \end{array} \end{equation}
Thus extrapolating the perturbative result for $\gamma$ to the non-perturbative regime leads to the BF bound  being violated when the perturbative running $\gamma =1/2$. Putting together (\ref{eq: anomalous_dimension}) and this factor of 2 we find a critical value of $\alpha$ for chiral symmetry breaking
\begin{equation}  \label{critc}
\alpha_c = \frac{\pi}{3~C_2(R)} ~.
\end{equation}

We can now perform a very simple computation, albeit improperly extending the one loop results beyond the perturbative regime. We assume that at a scale $\Lambda_{\chi SB}$ the SU(N) theory with $N_f(R)$ fermions in a given representation has $\alpha=\alpha_c$. We assume that at this scale the fermions will become massive and should be integrated from the IR theory below $\Lambda_{\chi SB}$. Now we are left in the IR with a pure glue theory which runs with $b_0 = 11 N/6 \pi$ so
\begin{equation} \alpha(Q^2) = {\alpha_c \over 1 + b_0^{\rm glue} \alpha_c \log(Q/\Lambda_\chi) } ~,\end{equation}
and the IR pole is then given by
\begin{equation} \Lambda_{\rm pole} = \Lambda_{\chi SB}{\rm exp}\left[{1 \over b_0^{\rm glue} \alpha_c} \right] ~.\end{equation}
We can combine these equations, to obtain and expression for the ratio of the two scales
\begin{equation} {\cal R}(R) =  { \Lambda_{\chi SB} \over \Lambda_{\rm pole} }= {\rm exp} \left( {9 \over 11} {N^2-1 \over N} {T(R) \over d(R)} \right) ~.\end{equation}

We will use this ${\cal R}(R)$ as a measure of the gap between the confinement scale and the chiral symmetry breaking scale. It is crude because: the perturbative results may not be a good description of the non-perturbative regime; confinement may happen before the Yang Mills theory pole (lowering the gap); the fermions may not sharply decouple from the running (potentially increasing the gap between the scales); and at strong coupling the two phenomena might become intertwined triggering each other. Nevertheless it stands as a straw man that can be used to ask questions on precisely these points.

\subsubsection{Conformal Window Constraint}

There is one additional constraint on the maximum number of $N_f(R)$ that we will note. It has been suggested \cite{Appelquist:1996dq, Ryttov:2007cx} that some of these theories have IR fixed points and if the fixed point value lies below $\alpha_c$ then they live in the ``conformal window'' and will not break chiral symmetry, nor confine.  For example the two loop beta function result for $\alpha$ is \cite{Ryttov:2007cx}
\begin{equation}
\mu \frac{d \alpha}{d \mu} = - b_0  \alpha^2 - b_1  \alpha^3  ~,
\end{equation} 
with 
\begin{equation} \label{eq: beta_function2}
\begin{aligned}
b_1 = \frac{1}{24 \pi^2} \left(34 C^2_{2}(G) - \sum_{R} (20 C_{2}(G) +  \right. \\
    \left.12 C_{2}(R)) T(R) N(R)  \vphantom{\frac{1}{2}} \right)~.
\end{aligned}
\end{equation}
$\beta$  vanishes at the fixed point so when
\begin{equation} \label{fpvalue}
\alpha_{\star} = -  ~ \frac{b_0}{b_1}~.
\end{equation}

Now one can compare this to the value of $\alpha_c$. 
We now find a new  upper limit on the number of flavours (lower than that at which asymptotic feedom is lost) given by
\begin{equation}
N^{\star}_{f}(R)_{\rm max} = \frac{d(G)}{d(R)} ~ \frac{C_{2}(G)}{C_{2}(R)} ~ \frac{17~C_{2}(G)+66~C_{2}(R)}{10~C_{2}(G) + 30~C_{2}(R)}~.
\end{equation} \newpage

\subsubsection{Results and Outlook for One Representation SU$(N)$}

\begin{figure*}
    \includegraphics[width=\textwidth]{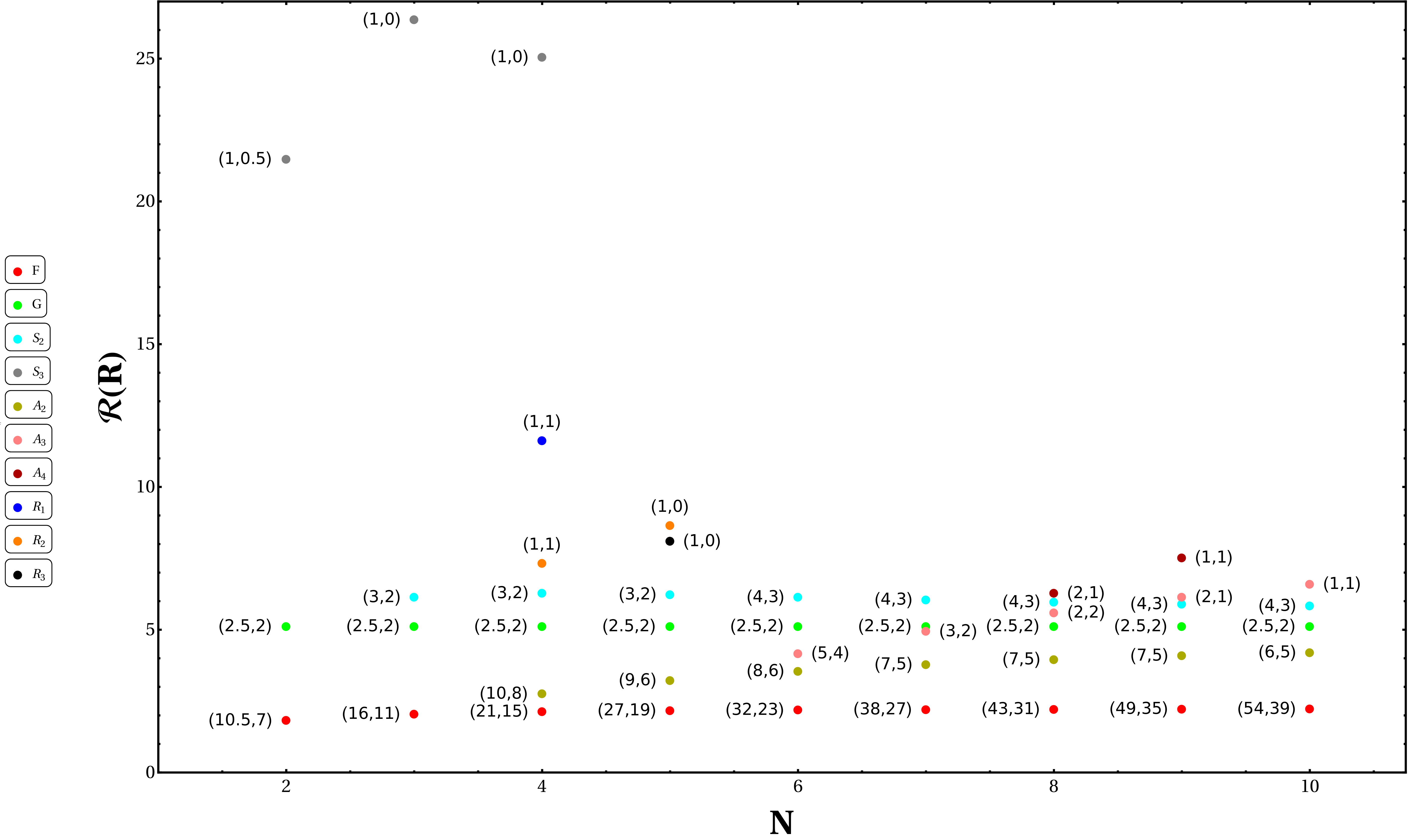}
\caption{Plot of ${\cal R}(R) =  {\Lambda_{\chi SB} / \Lambda_{\rm pole} }$ against $N$ for various single representation theories:
we have used red for the fundamental, green for the adjoint, cyan for the rank-2 symmetric, gray for three-rank symmetric, gold for the rank-2 antisymmetric, pink for the $A_3$, maroon for the $A_4$, and blue, orange, black for the R$_{1,2,3}$ respectively. The points are marked by the maximum number of flavours for which the theory is asymptotically free and the lower number of flavours that marks the last chiral symmetry breaking theory before the  conformal window begins.
}
\label{fig: plots_sun_1} 
\end{figure*}

In \Cref{fig: plots_sun_1}  we plot the ratio ${\cal R} = { \Lambda_{\rm pole} / \Lambda_{\chi SB} }$, which measures the split between the chiral symmetry breaking scale and the confinement scale, for all the asymptotically free theories containing the representations we have listed. The points are labelled by the (integer) maximum value of flavours $N_f(R)$ for the theory to be asymptotically free. For SU(2) where representations are real, and generically for the real adjoint representation, we allow 1/2 integer values of $N_f$. 
We also include in the label the maximum number of flavours for the theory to lie below the conformal window at the level of the two loop approximation - this is the second number associated with each point in \Cref{fig: plots_sun_1}. 

An important point to note here is that we include theories with $N_f=1,1/2$ where the axial symmetry is expected to be anomalous. The reason is that the anomaly is driven by the operator $Tr F \tilde{F}$  but this vacuum expectation value may be associated with the confinement scale. If there is a big gap between the confinement and chiral symmetry breaking scales then the anomaly may be a minor issue at the fermion condensation scale.

Note that in the large-$N$ limit we can explicitly compute the limiting value for the ratio of scales  for the 4 representations that remain asymptotically free at large $N$
 \begin{align}
\begin{split}
{\cal R}(F) &= e^{\frac{9}{11}}=2.27, \qquad {\cal R}(G) = e^{\frac{18}{11}} = 5.14\\
{\cal R}(S_2) &= e^{\frac{18}{11}}=5.14, \qquad {\cal R}(A_2) = e^{\frac{18}{11}}=5.14
\end{split}
\end{align}  

We will now review our results and relevant lattice simulation results. Note that the lattice effort to date has largely been concentrated on Beyond the Standard Model (BSM) physics and seeking walking gauge theories. This is neccesarily challenging on the lattice since these theories have an indistinct chiral symmetry breaking scale (the coupling runs slowly at $\alpha_c$) and there can be a large gap in scales between the weak coupling and chiral symmetry breaking scale.  Whether a theory lives in the conformal window or breaks chiral symmetry at a low scale is often hard to determine so many questions remain to be resolved.\bigskip

\makeatletter\onecolumngrid@push\makeatother
\clearpage
\makeatletter\onecolumngrid@pop\makeatother

\noindent {\it Fundamental Matter:} To begin to place the plot in \Cref{fig: plots_sun_1} in context let us begin by considering the case of fundamental representation fermions. Here the plot has ${\cal R}\simeq2$ in all cases. However, we know of no lattice results that have claimed to see a clear distinction between the chiral and deconfinement transition at finite temperature (see for example \cite{Aoki:2006br, Karsch:2001cy}). We must therefore take ${\cal R}=2$
to represent no observable gap. 

There has been considerable work on studying SU(3) theories with varying numbers of fundamental fields. The $N_f=8$ theory is known to likely break chiral symmetries \cite{Appelquist:2007hu, Appelquist:2014zsa, Appelquist:2018yqe}. The $N_f=12$ theory is likely in the conformal window (with $g_*^2=6.2(2)$ \cite{Appelquist:2012nz, Lin:2015zpa, Fodor:2016zil,  Cheng:2014jba}) . Recent work \cite{Hasenfratz:2020ess} has centred on testing $N_f=10$ with hints that it lies in the conformal window or has a very slow IR running. The two loop prediction that the critical value of $N_f$ is 11 is still a reasonable estimate. 
SU(2) theories have also been studied - see \cite{Karavirta:2011zg}. Here signs of IR fixed point behaviour are observed above $N_f=6$ so the two loop prediction for the edge of the conformal window may be a little high. \bigskip

\noindent {\it Adjoint Matter:} for which ${\cal R}=5$ for all $N$ might be the first case on our plot when we begin to have some confidence that an observable gap could be found. The SU(3) theory with two Dirac flavours has been studied on the lattice in \cite{Karsch:1998qj} and there a gap size of a factor of 8 is found between the confinement and chiral symmetry breaking scales (the former is a first order transition, whilst the latter is continuous). That study provides strong support for the hypothesis that the two phenomena are separated.
That the gap size is larger than we predict is also possibly evidence that at strong coupling the fermions do not decouple sharply from the running below their mass scale, but continue to slow the running to the IR Yang Mills theory pole.  This should be caveated by the possibility the theory is rather walking above the chiral symmetry breaking scale \cite{DeGrand:2013uha}  which makes lattice simulations hard - \cite{Karsch:1998qj} may not have corresponded to the continuum limit.

SU(2) theories with $N_f$=2 \cite{DelDebbio:2008zf, DeGrand:2011qd, Rantaharju:2015yva} and $N_f$=1 \cite{Athenodorou:2014eua} have also been simulated each showing some signs of fixed point beahviour but as yet no concrete obeservation of chiral symmetry breaking has been reported. 

We also note that the case of a single Weyl adjoint degree of freedom is the ${\cal N}=1$ super Yang-Mills theory which has also been studied on the lattice \cite{Bergner:2014saa} - here the second order transition with temperature for both the Polyakov loop and the chiral condensate appear to occur at the same scale (within the errors). That this case is different is not surprising because the supersymmetry ties glueballs and gluino balls into the same supersymmetric multiplets \cite{Veneziano:1982ah} so the supersymmetry will naturally bring the two scales together, physics our computations totally miss. 
\bigskip

\noindent {\it Two Index Symmetric Matter:} here we predict a potentially observable gap between confinement and chiral symmetry breaking as for the adjoint matter fields. There has been considerable interest in the SU(3) $N_f=2$ theory since it might be rather walking above the chiral symmetry breaking scale. In \cite{DeGrand:2008kx} with Wilson fermions a gap between confinement and chiral symmetry breaking  was not observed. On the other hand in \cite{Kogut:2010cz} with staggered fermions a gap  was observed.  Work since these papers \cite{Kogut:2015zta, Fodor:2015zna, Hansen:2017zyo} has centred on determining how walking the dynamics is with sufficient suggestion of slow running  to make lattice results hard to confidently interpret at this stage. The focus to date has been on looking for walking dynamics but it might be interesting to study for example SU(6) with $N_f=2$ which is likely away from the edge of the conformal window. That theory might clarify the size of the gap if one is interested in the confinement versus chiral symmetry breaking separation. \bigskip

\noindent {\it More exotic representations:} We also have predictions for representations that have not been studied yet on the lattice. The $R_2$ and $R_3$ representations are only asymptotically free for particular choices of $N$ and then likely in the conformal window. The predictions here would be for gaps between confinement and chiral symmetry breaking like that for the adjoint representation. The SU(4) $N_f=1$, $R_1$ theory is of more interest since that theory is predicted to have a gap of over 11 which could be straightforward to see on the lattice at finite temperature where one might observe chiral symmetry breaking but no sign of confinement even at quite low temperatures. 
Finally the $S_3$ theories stand out for having gaps above 20. Only the SU(2) theory with a single Weyl fermion is likely outside the conformal window though. That theory where the $S_3$ is only 4 dimensional might though also be of interest to study. \bigskip

The interesting theories with large gaps between confinement and chiral symmetry breaking are likely (or have already proven) hard to study in lattice simulations because of the potentially slow running above the chiral symmetry breaking scale and the closeness of the critical coupling to the fixed point value. For example consider the following data for four of the theories \bigskip

\begin{tabular}{llll}
 SU(3), $F$, $N_f=3$ & $b_0=1.43$    &  $\alpha_c =0.79$   & $\alpha_*=\infty$ \\
&&&\\
SU(3), $G$, $N_f=2$ & $b_0=0.48$    &  $\alpha_c =0.35$   & $\alpha_*=0.42$ \\
&&&\\
SU(4), $R_1$, $N_f=1$  & $b_0=0.64$    &  $\alpha_c =0.17$   &  $\alpha_*=0.22$ \\
&&&\\
SU(2), $S_3$, $N_f=0.5$  & $b_0=0.64$    &  $\alpha_c =0.28$   &  $\alpha_*=1.97$ \\
\end{tabular} \bigskip

The first is QCD and has fast running from the perturbative regime (large $b_0$) and a critical coupling below the fixed point value (which is formally infinite). It has a very clear distinction in phase between the perturbative regime and the strong coupling regime and the chiral symmetry breaking scale is associated with fast running so clear cut. The SU(3) model with adjoints and the SU(4) model with $R_1$ both run more slowly from the perturbative regime but crucially have the critical coupling very close to the fixed point value making the chiral symmetry breaking scale indistinct. This has actually been the motivation to study the adjoint theory but neccesarily makes the job on the lattice extremely hard. From the perspective of just observing the gap between confinement and chiral symmetry breaking the SU(2) theory with an $S_3$ looks much easier to study since the critical coupling lies quite below the fixed point value so the chiral symmetry breaking scale should be easier to identify.

\section{Two-representation theories}

In this section we will move to considering gauge theories with the matter fields distributed in two distinct and inequivalent representations of the gauge group.
Such theories are starting to become of interest because they may play a role in composite higgs \cite{Ferretti:2013kya}, composite dark matter \cite{Sannino:2009za} and composite inflation \cite{Channuie:2011rq} models. As a result there are already some lattice simulation of such theories. Here our interest began as whether we could construct a model where a higher dimension representation condenses at one scale leaving a conformal window theory below the scale where the higher representation was integrated out. Such a theory essentially would never confine although it would have chiral symmetry breaking. In fact at the level of the approximations we use we have not found any examples of this behaviour. 

In spite of this one can achieve theories where a higher dimension representation condenses leaving a theory with a somewhat slow running coupling at lower scales that generates a sizable gap before the lower dimensional representation condenses and then presumably confinement sets in. We will concentrate on theories where the lowest dimension representation is the fundamental  since every added flavour here has the smallest possible change on the beta function coefficients, giving the most ability to tune the running. Further since we want to be able to add as many as we possibly can to bring the theory with just  fundamentals towards the conformal window, we add the minimum number of flavours of the higher dimension representation. That is one Dirac spinor for complex representations and one Weyl spinor for real representations (we label  this as  $1/2$ a Dirac spinor in plots) - in fact we will also show results for 1 Dirac flavour in the case of the adjoint representation to highlight the difference.

We proceed to perform another straightforward computation based on the perturbative running results in a gauge theory with $N_f$  fields in the fundamental representation plus one (or a 1/2) in the higher representations from the previous section. Again here we assume that if the confinement scale lies far below the chiral symmetry breaking scale then we can neglect the axial anomaly. We compute $b_0$ (\ref{eq: beta_function1}) and $b_1$  (\ref{eq: beta_function2}) in the theory with both representations present to check the theory is asymptotically free, but 
also that the IR fixed point value (\ref{fpvalue}) lies above the critical coupling  for the higher dimension representation $\alpha_c^R$ (\ref{critc}). We then ask 
 what is the maximum value of $N_f$ such that $\alpha_* > \alpha_c^R$. In that theory we assume that at some scale $\Lambda_{\chi SB ~ R}$ the coupling has run equal to $\alpha_c^R$ and the heavy fermions are 
 integrated out. Next we run the coupling numerically into the IR for the theory with just the (maximal number of $N_f$) fundamentals. We ask at what scale, $\Lambda_{\chi SB ~F}$  it reaches the critical coupling for the fundamental fields.

The ratio of these two scales which we denote by $\mathcal{Q}(R)$ 
\begin{equation} {\mathcal Q}(R) = {\Lambda_{\chi SB ~ R} \over \Lambda_{\chi SB ~F} } \end{equation}
is the gap between the two condensation scales for the given representation $R$. Since we expect the confinement scale to lie below $\Lambda_{\chi SB ~F}$, this also measures the gap between the chiral symmetry breaking scale for $R$ and the confinement scale.  Note that the gap to the pole of the theory is given by ${\mathcal R}(R) {\mathcal Q}(R)$ and using the results of  \Cref{fig: plots_sun_1}  the gap to confinement could be bigger than just ${\mathcal Q}(R)$.

\begin{figure*}
    \includegraphics[width=\textwidth]{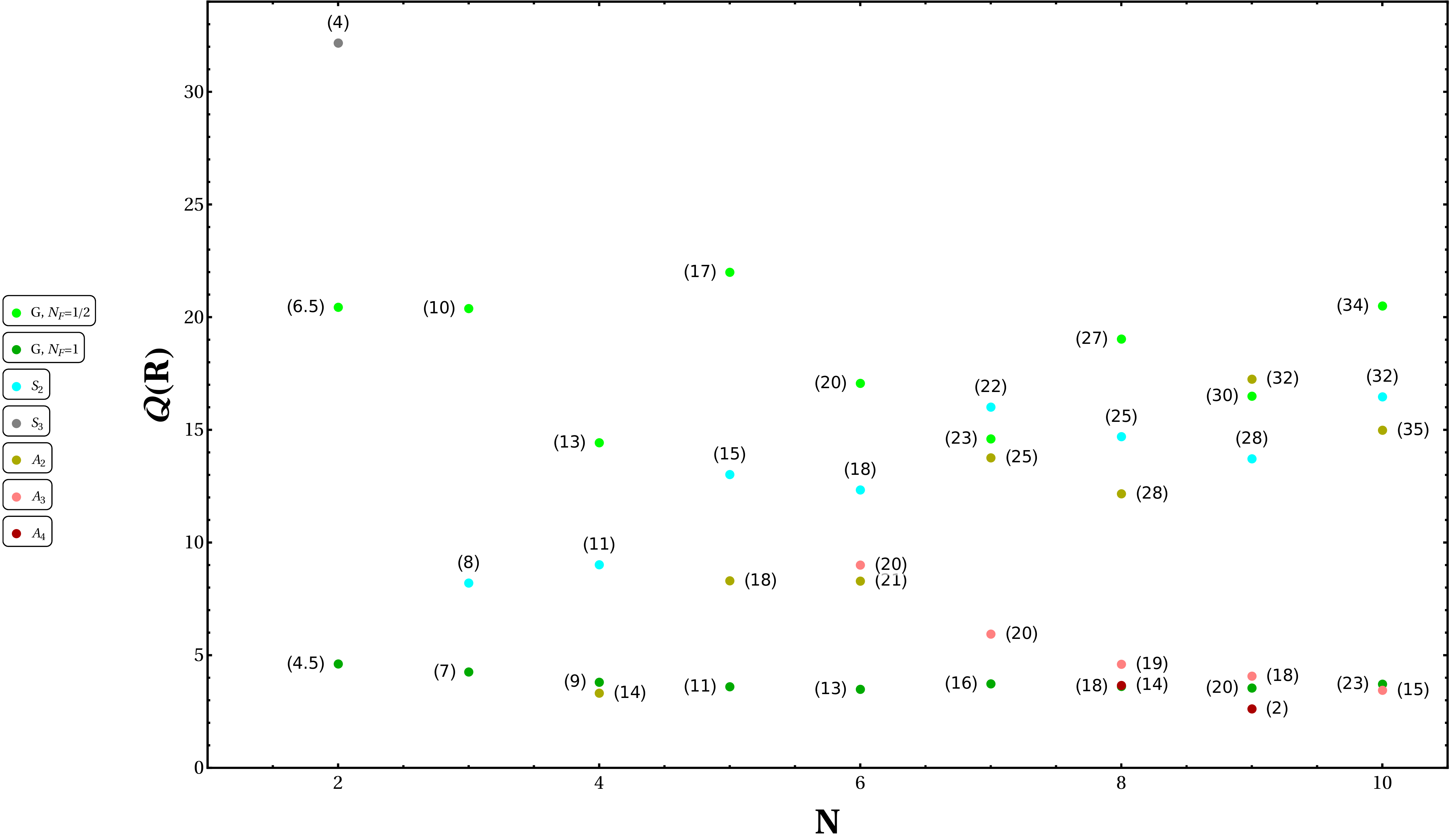}
\caption{
Plot of $ {\mathcal Q}(R) = {\Lambda_{\chi SB ~ R} / \Lambda_{\chi SB ~F} }$ against $N$ in theories with the minimal number of fermions in the higher dimensional representation (either 1 or 1/2 for real representations) and $N_f^F$ in the fundamental representation. $N_f^F$ has been tuned to maximize $\mathcal{Q}(R)$ and its value is next to the point. We have used red for the fundamental, green for the adjoint, cyan for the rank-2 symmetric, gray for three-rank symmetric, gold for the rank-2 antisymmetric, pink for the $A_3$, maroon for the $A_4$, and blue, orange, black for the R$_{1,2,3}$ respectively.
}
\label{fig: plots_sun_2} 
\end{figure*}

We present our results in \Cref{fig: plots_sun_2}, where we display the maximum value of ${\mathcal Q}(R)$ we can find by varying $N_f^F$ as a function of $N$ for each possible representation $R$. We label the points by the number of Dirac fermions in the fundamental representation which has been used to maximize the gap.

One immediately sees that there are many theories with adjoint, $S_2$ or $A_2$ representations that have gaps in excess of a factor of ten. Adding four fundamentals to the SU(2) theory with an $S_3$ raises the gap to over a factor of 30. The convincing discovery of such a gap in a lattice simulation would certainly show confinement and chiral symmetry breaking to be totally separate phenomena.

We must be careful though because by tuning the gap large we are also potentially making life harder for lattice simulations. As an example lets consider SU(3) with a single Weyl fermion in the adjoint. This is just ${\cal N}$=1 super Yang Mills. Now we can consider adding fundamental fermions (which breaks supersymmetry) to observe the gap growing - here our $b_0$, $\alpha_c$ and $\alpha_*$ are for the theory with both representations present above the first chiral symmetry breaking transition for the adjoint: \medskip

\begin{tabular}{lllll}
$N_f^F$=0 & $b_0$=1.43    &  $\alpha_c$=0.35   & $\alpha_*$=$\infty$ & \\
&&&\\
$N_f^F$=4 & $b_0$=1.01    &  $\alpha_c$=0.35   & $\alpha_*$=$\infty$ & ${\Lambda_{\chi SB ~ R} \over \Lambda_{\chi SB ~F} }$= 2.6\\
&&&\\
$N_f^F$=8  & $b_0$=0.58    &  $\alpha_c$=0.35   &  $\alpha_*$=0.97 & ${\Lambda_{\chi SB ~ R} \over \Lambda_{\chi SB ~F} }$=5.8\\
&&&\\
$N_f^F$=10  & $b_0$=0.37    &  $\alpha_c$=0.35   &  $\alpha_*$=0.40 & ${\Lambda_{\chi SB ~ R} \over \Lambda_{\chi SB ~F} }$=20.3\\
\end{tabular} \medskip

The $N_f^F=0$ theory is QCD-like with fast running (large $b_0$) and $\alpha_c \ll \alpha_*$. As we add in fundamental fields we slow the running ($b_0$ decreases) and $\alpha_*$ falls, as the gap between chiral symmetry breaking for the two representations widens. The $N_f^F=10$ theory has $\alpha_c$ very 
\makeatletter\onecolumngrid@push\makeatother
\makeatletter\onecolumngrid@pop\makeatother

\noindent  close to $\alpha_*$ and the lattice will most likely struggle to identify the point. The $N_f^F=8$ theory might represent 
 a compromise 
  that allows the separation to be seen more cleanly even though the gap is smaller. Incidentally $N_f^F=8$ can be implemented with staggered fermions so would also be cheaper (the single adjoint field would need more sophisticated methods). Further it has been identified as lying outside the conformal window on the lattice already \cite{Appelquist:2007hu, Appelquist:2014zsa, Appelquist:2018yqe}.

In a similar vein it is probably not sensible to add fundamental fields to the SU(2) theory with a Weyl $S_3$ since the gap is predicted to be large already so adding in walking behaviour will only complicate the simulations. 

Finally we note a number of other promising candidate theories with large gaps where fundamental fields could be included as staggered fermions, albeit at larger $N$ values: \medskip

\begin{tabular}{l}
$\{ SU(5) ~ | ~ 16~F, 1/2 G \}$ with $\mathcal{Q}=12.2$ \\
$\{ SU(9) ~ | ~ 28~F, 1/2 G \}$ with $\mathcal{Q}=9.55$ \\
$\{ SU(10) ~ | ~ 32~F, 1/2 G \}$ with $\mathcal{Q}=11.5$ \\
$\{ SU(7) ~ | ~ 20~F, 1 S_2 \}$ with $\mathcal{Q}=9.24$ \\
$\{ SU(8) ~ | ~ 24~F, 1 S_2 \}$ with $\mathcal{Q}=11.3$
\end{tabular}

\subsubsection{Two Representation Lattice Studies}

There have already been a number of lattice studies of SU($N$) theories with two representations. In \cite{Ayyar:2018ppa} SU(4) with two F and two $A_2$ has been studied and a single deconfinement and chiral symmetry restoration transition observed (it is first order). This is not surprising given that $N_f^F=2$ is low and the theory lies close to the pure $A_2$ theory running. Here we do not expect really a bigger gap that in QCD (see \Cref{fig: plots_sun_1}).

In \cite{Engels:2005rr} the SU(3) theory with adjoints was supplemented by $N_f^F=2$ fundamentals where a gap between chiral symmetry breaking and confinement was again seen as in the $N_f^F=0$ model \cite{Karsch:1998qj} (again care may be needed to find the continuum limit). This does not push $N_f^F$ up as high as 10 as we have suggested to maximise the gap but shows the lattice technology does exist to study such theories. 

Very recently \cite{Bergner:2020mwl} has begun a study  of theories with a Weyl adjoint and fundamentals. For $N_f^F=2$ the theory has been identified as breaking chiral symmetries but the temperature phase structure has not yet been explored.

\section{Conclusions}

We have reviewed old arguments that chiral symmetry breaking and confinement may be distinct phenomena that are just accidentally close in scale for QCD. We have presented some simple computations based on the two loop running results for $\alpha$ and $\gamma$ for gauge theories with higher dimensional representations. We have sought theories with one representation with the largest possible gap between the scale where $\gamma=1$ (and chiral symmetry breaking occurs) and the pole of the running in the deep IR pure glue theory where confinement might be associated. We have found example theories with much larger gaps than QCD. This view is supported by the work in \cite{Karsch:1998qj} which shows such a gap for adjoint matter. 

We have also highlighted how including additional matter in the fundamental representation can slow the running between the higher representation's chiral symmetry breaking scale and that of the fundamental representation. The confinement scale should then be a little lower yet, so this is further evidence that chiral symmetry breaking and confinement are disconnected. These theories provide an alternative test of the idea of walking in gauge theories, as well as adding further evidence of the gap between confinement and chiral symmetry breaking. 

The plots we have presented, based as they are on perturbative results extended to the non-perturbative regime, are only representative of the expected phenomena. It would be very interesting to study such theories on the lattice to provide first principle confirmation of the ideas. 
Lattice methods now extend to theories with multiple and higher representations and have concentrated on looking at walking theories that might apply to Beyond the Standard Model (BSM) physics. We think it would be sensible to study also the separation of chiral symmetry breaking and confinement  because it will shed light on the fundamental mechanisms at work at strong coupling and reinforce the logic behind the BSM theories. These models we propose are also less close to the conformal limit than walking theories so should be easier to simulate on the lattice. 
In the case of two representation theories,  by having a sharp UV scale where the higher dimension representation  condenses before the walking regime in the theory with just fundamental fields begins, it may help lattice simulations more cleanly understand walking dynamics.  

Note we have assumed in our discussion that finite temperature transitions will occur in sequence as the temperature passes through the scale of each phenomena as seen at zero temperature. For example, as one raises the temperature, that first the theory will deconfine then sequential representations of fermions will have chiral symmetry restored. We assume gaps between these temperatures would reflect the gaps between the scales we have computed. However, at finite temperature, 
thermal zero modes of the gauge degrees of freedom may control the dynamics
towards the infrared whereas the fermions decouple as they acquire a thermal
“Matsubara mass”. This may change the dynamics compared to the zero-temperature
case considered in the present work. It would therefore be good to confirm this interplay in a first principle lattice calculation. 

In the introduction we noted that NJL operators have already been used to separate the confinement and chiral symmetry breaking scales \cite{Sinclair:2008du}. In these theories it would be very interesting to follow the behaviour of the Polyakov loop as the chiral symmetry breaking scale is raised. If operator mixing does tend to bind the two scales together then one would expect the confinement scale at first to grow with the NJL coupling. Eventually it would be expected to fall lower as the gauge coupling becomes weakly coupled at the chiral symmetry breaking scale. Such an analysis could shed light on a critical coupling for confinement. An interesting renormalization group analysis of these theories has been performed in \cite{Braun:2012zq} and supports the idea that the scales can be separated in the presence of large NJL operators in theories with fundamental and adjoint quarks. That study also suggests that at low NJL coupling there may be some binding of the two scales.

Finally, if the separation between confinement and chiral symmetry breaking is well established it may have consequences for QCD where at high density new phases such as deconfined massive phases \cite{Fadafa:2019euu,BitaghsirFadafan:2020otb} or quarkyonic phases \cite{McLerran:2007qj} may exist if such separation is allowed.

\section*{Acknowledgments}~ We thank Kimmo Tuominen, Biagio Lucini and Georg Bergner for discussions. NEs work was supported by the STFC consolidated grant ST/P000711/1.

\bibliographystyle{apsrev4-1} 
\bibliography{lit.bib}

\end{document}